\documentclass{elsart}
\usepackage{graphicx}

\def\l{\langle}
\def\r{\rangle}

\begin{document}
\begin{frontmatter}
\title{
Novel Monte Carlo algorithms \\
and their applications
}

\author{Yutaka Okabe\thanksref{okabe}},
\author{Yusuke Tomita},
\author{Chiaki Yamaguchi}

\address{
Department of Physics, Tokyo Metropolitan University,
Hachioji, Tokyo 192-0397, Japan
}

\thanks[okabe] {Electronic address: okabe@phys.metro-u.ac.jp}

\date{}

\begin{abstract} 
We describe a generalized scheme for the probability-changing cluster 
(PCC) algorithm, based on the study of the finite-size scaling 
property of the correlation ratio, 
the ratio of the correlation functions with different distances. 
We apply this generalized PCC algorithm to 
the two-dimensional 6-state clock model.  
We also discuss the combination of the cluster algorithm and 
the extended ensemble method.  We derive a rigorous broad histogram 
relation for the bond number.  A Monte Carlo dynamics 
based on the number of potential moves for the bond number is proposed, 
and applied to the three-dimensional Ising and 3-state Potts models.  
\end{abstract}


\end{frontmatter}

\section{Introduction}

In the Monte Carlo simulation, we sometimes suffer from 
the problem of slow dynamics.  
The critical slowing down near the critical point, 
the phase separation dynamics at low temperature, 
the slow dynamics due to the randomness or frustration, 
and the low-temperature slow dynamics in quantum Monte Carlo simulation 
are examples of the problems of slow dynamics.  

We may classify the attempts to conquer the slow dynamics in
the Monte Carlo simulation into two categories.  
The first category is the cluster algorithm, such as
the Swendsen-Wang (SW) algorithm \cite{sw87} and the Wolff 
algorithm \cite{wolff89}.  
The second one is the extended ensemble method. \
The multicanonical method \cite{berg91,Lee93}, 
the broad histogram method \cite{oliveira96}, 
and the flat histogram method \cite{wang98,Wang00} are examples 
of the second category.  Recently, Wang and Landau \cite{wl01} 
proposed an efficient algorithm to accelerate the calculation of 
the energy density of states (DOS). 
Yamaguchi and Okabe \cite{yo01} have successfully used 
the Wang-Landau algorithm for the study of the antiferromagnetic 
$q$-state Potts models.

Tomita and Okabe \cite{PCC} recently proposed 
an effective cluster algorithm, which is called the 
probability-changing cluster (PCC) algorithm, 
of tuning the critical point automatically.  
The PCC algorithm is an extension of the SW algorithm \cite{sw87}; 
we change the probability of connecting spins of the same type, 
essentially the temperature, 
in the process of the Monte Carlo spin update.   
We showed the effectiveness of the PCC algorithm 
for the two-dimensional (2D) and three-dimensional (3D) Potts 
models \cite{PCC}, determining the critical point and exponents. 
We can extract information on critical phenomena with much 
less numerical effort.  
The PCC algorithm is quite useful for studying random 
systems, where the distribution of the critical temperature, $T_c$, 
is important. 
We applied the PCC algorithm to the 2D 
diluted Ising model \cite{to01b}, investigating the crossover 
and self-averaging properties of random systems. 
We also extended the PCC algorithm to the problem of 
the vector order parameter \cite{XY}; studying the 2D 
XY and clock models, we showed that the PCC algorithm 
is also useful for the Kosterlitz-Thouless (KT) transition \cite{KT}.

The combination of approaches of two categories, 
the cluster algorithm and the extended ensemble method, is a challenging 
problem. 
Janke and Kappler \cite{Janke} proposed a trial to combine 
the multicanonical method and the cluster algorithm; 
their method is called the multibondic ensemble method. 
Recently, Yamaguchi and Kawashima \cite{Yama02} 
improved the multibondic ensemble method, and also showed 
that the combination of the Wang-Landau algorithm and 
the improved multibondic ensemble method yields much better statistics 
compared to the original multibondic ensemble method \cite{Janke}.  

Here, we pick up two recent topics of new Monte Carlo 
algorithms.  We first discuss the generalization 
of the PCC algorithm.  This generalized scheme is 
based on the finite-size scaling (FSS) 
property of the correlation ratio. 
Second, for the algorithm to combine the cluster algorithm 
and the extended ensemble method, we derive a rigorous broad 
histogram relation for the bond number, and propose  
the flat histogram method for the bond number.

\section{Generalized scheme for the PCC algorithm}

We start with reviewing the idea of the PCC 
algorithm \cite{PCC}.  
We explain the case of the ferromagnetic $q$-state Potts model, 
whose Hamiltonian is given by
\begin{equation}
  \mathcal{H} = -J \sum_{<i,j>} (\delta_{\sigma_i,\sigma_j}-1), \quad 
  \sigma_i = 1, 2, \cdots, q.
\label{Hamiltonian}
\end{equation}
We construct the Kasteleyn and Fortuin (KF) \cite{KF} clusters using the probability $p$, 
where $p$ is the probability of connecting spins of the same type, 
$p = 1 - e^{-J/k_BT}$.  
The correspondence of the spontaneous magnetization of the $q$-state 
Potts model and the percolation probability of the bond 
percolation model was discussed by Hu \cite{Hu84}. 
Then, we check whether the system is percolating or not. 
If the system is percolating (not percolating) in the previous test, 
we decrease (increase) $p$ by $\Delta p \ (>0)$.
Spins are updated following the same rule as the SW algorithm. 
After repeating the above processes, the distribution of $p$ 
for Monte Carlo samples approaches the Gaussian distribution 
of which mean value is $p_c(L)$;  $p_c(L)$ is the probability 
of connecting spins, such that the existence probability 
$E_p$ becomes 1/2.  
The existence probability $E_p$ \cite{Hu92} 
is the probability that the system percolates. 
Since $E_p$ follows the FSS near the critical point, 
\begin{equation}
  E_p(p,L) \sim X(t L^{1/\nu}), \quad t=(p_c-p)/p_c ,
\label{scale}
\end{equation}
where $p_c$ is the critical value of $p$ for the infinite system 
and $\nu$ is the correlation-length 
critical exponent,
we can estimate $p_c$ from the size dependence of $p_c(L)$ 
using Eq.~(\ref{scale}) and, in turn, estimate $T_c$ 
through the relation $p_c = 1 - e^{- J/ k_BT_c}$.

In the original formulation of the PCC algorithm \cite{PCC}, 
we used the KF representation 
in two ways.  First, we make a cluster flip as 
in the SW algorithm \cite{sw87}. 
Second, we change the probability of connecting spins of 
the same type, $p$, depending on the observation 
whether clusters are percolating or not. 
The point is that $E_p$ has 
the FSS property with a single scaling variable. 
We may use quantities other than $E_p$ which have 
a similar FSS relation.
In the FSS analysis of the Monte Carlo simulation, 
we often use the Binder ratio \cite{Binder}, which is 
essentially the ratio of the moments of the order parameter $m$. 
The moment ratio has the FSS property with a single scaling variable,
\begin{equation}
  \frac{\l m^4 \r}{\l m^2 \r^2} = f(tL^{1/\nu}),
\label{mom_ratio}
\end{equation}
as far as the corrections to FSS are negligible.  The angular 
brackets indicate the thermal average.  
The moment ratio derived from a snapshot spin configuration 
is always one; therefore, the instantaneous moment ratio 
cannot be used for the criterion 
of judgment whether we increase or decrease the temperature. 

As another quantity, we may treat the correlation ratio, 
the ratio of the correlation functions 
with different distances. 
For an infinite system at the critical point, the correlation 
function decays as a power of $r$, 
\begin{equation}
  \l g(r) \r \sim r^{-(D-2+\eta)}, \quad (L=\infty, \ t=0),
\label{g(r)}
\end{equation}
with the decay exponent $\eta$. 
Precisely, the distance $r$ is 
a vector, but we have used a simplified notation.  
Away from the critical point, the ratio of the distance $r$ 
and the correlation length $\xi$ plays a role 
in the scaling of the correlation function. 
Moreover, for finite systems, two length ratios, $r/L$ and 
$L/\xi$, come in 
the scaling form of the correlation function.
Then, the ratio of the spin-spin correlation functions with 
different distances $r$ and $r'$ 
takes the FSS form with a single scaling variable, 
\begin{equation}
 \frac{\l g(r,t,L) \r}{\l g(r',t,L) \r}= \tilde f(L/\xi),
\label{corr_ratio}
\end{equation}
if we fix two ratios, $r/L$ and $r/r'$. 
In case the correlation length diverges with a power law, 
$\xi \propto t^{-\nu}$, Eq.~(\ref{corr_ratio}) becomes 
the same form as Eq.~(\ref{mom_ratio}); however, 
Eq.~(\ref{corr_ratio}) is also applicable 
to the case of the KT transition, where the correlation 
length diverges more strongly than the power-law divergence. 

\begin{figure}[th]
\begin{center}
\includegraphics[width=.5\textwidth]{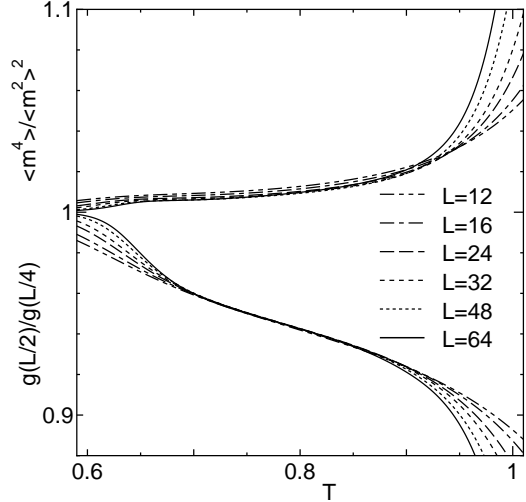}
\end{center}
\caption[]{
The moment ratio $\l m^4 \r/\l m^2 \r^2$ and 
the correlation ratio $\l g(L/2) \r/\l g(L/4) \r$
of the 2D 6-state clock model for $L$ = 12, 16, 24, 32, 48, and 64. 
} 
\label{fig_1}
\end{figure}

In order to examine the FSS properties of the correlation ratio, 
we here give the result of the 2D 6-state clock model 
on the square lattice with the periodic boundary conditions. 
The 2D $q$-state clock model is known to exhibit 
two phase transitions of the KT type at $T_1$ and $T_2$ 
($T_1<T_2$) for $q>4$ \cite{Jose}. 
We simulate the 2D 6-state clock model 
by the use of the Wang-Landau algorithm \cite{wl01}. 
We show the temperature dependence of both 
the moment ratio $\l m^4 \r/\l m^2 \r^2$ 
and the correlation ratio $\l g(L/2) \r/\l g(L/4) \r$ 
in Fig.~\ref{fig_1}. 
As for the distances $r$ and $r'$, we choose $L/2$ and $L/4$; 
we take the horizontal or vertical direction of the lattice 
for the orientation of two sites. 
For the correlation ratio, the curves 
of different sizes merge in the intermediate KT phase 
($T_1<T<T_2$), and spray out for the low-temperature ordered 
and high-temperature disordered phases, which is expected from 
the FSS form of Eq.~(\ref{corr_ratio}). 
Then, we can make a FSS analysis based on the KT form of 
the correlation length, 
\begin{equation}
 \xi \propto \exp(c/\sqrt{t}),
\label{corr_length} 
\end{equation}
where $t=|T-T_{\rm KT}|/T_{\rm KT}$.  
Using the data of $\l g(L/2) \r/\l g(L/4) \r$ 
for $L$=12, 16, 24, 32, 48, and 64, 
we estimate two KT transition temperatures. 
The best-fitted estimates are 
$T_1$=0.698(4) and $T_2$=0.898(4), in units of $J/k_B$, 
which are compatible 
with the recent results using the PCC algorithm \cite{XY}, 
$T_1$=0.7014(11) and $T_2$=0.9008(6). 
On the contrary, as seen from Fig.~\ref{fig_1}, 
the corrections to FSS are larger for the moment ratio, 
which makes the FSS analysis difficult. 
We have shown that the correlation ratio is a good 
estimator especially for the KT transition. 
It is due to the fact that we only use the property of 
correlation function; the characteristic of the KT transition 
is that the correlation function shows a power-law decay 
at all the temperatures of the KT phase.

We may use the FSS properties of the correlation ratio for 
the generalization of the PCC algorithm.  
Instead of checking whether the clusters are percolating or not, 
we ask whether the instantaneous correlation ratio $g(L/2)/g(L/4)$ 
is larger or smaller than some fixed value $R_c$. 
Of course, we can use other sets of distances. 
We decrease (increase) the temperature, if $g(L/2)/g(L/4)$ 
is smaller (larger) than $R_c$. 
We start the simulation with some temperature. 
We make the amount of the change of temperature, $\Delta T$, 
smaller during the simulation; in the limit of $\Delta T \to 0$, 
the system approaches the canonical ensemble.  

As an example, we apply the generalized scheme of the PCC algorithm 
to the study of the 2D 6-state clock model.  
We treat the systems with linear sizes $L$ = 8, 16, 32, 
64, and 128.  We start with $\Delta T$ = 0.005, 
and gradually decrease $\Delta T$ to the final value, 0.0001. 
After 20,000 Monte Carlo sweeps of determining $T_{\rm KT}(L)$, 
we make 10,000 Monte Carlo sweeps to take thermal average; 
we make 20 runs for each size to get better statistics 
and to evaluate the statistical errors. 
We calculate $g(L/2)/g(L/4)$ to check whether it is 
larger than $R_c$ or not. 

\begin{figure}[th]
\begin{center}
\includegraphics[width=.48\textwidth]{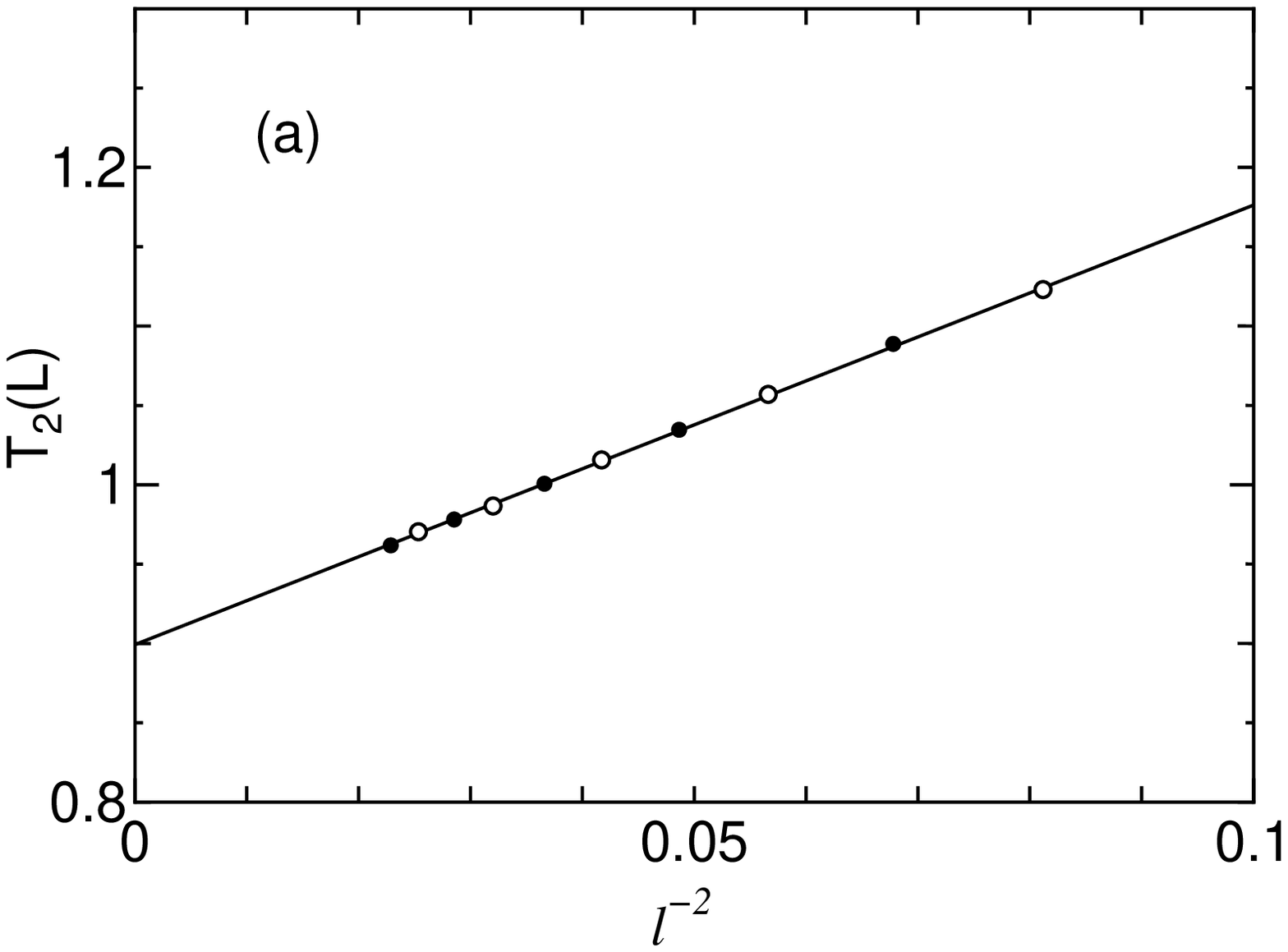}
\hspace{0.02\textwidth}
\includegraphics[width=.48\textwidth]{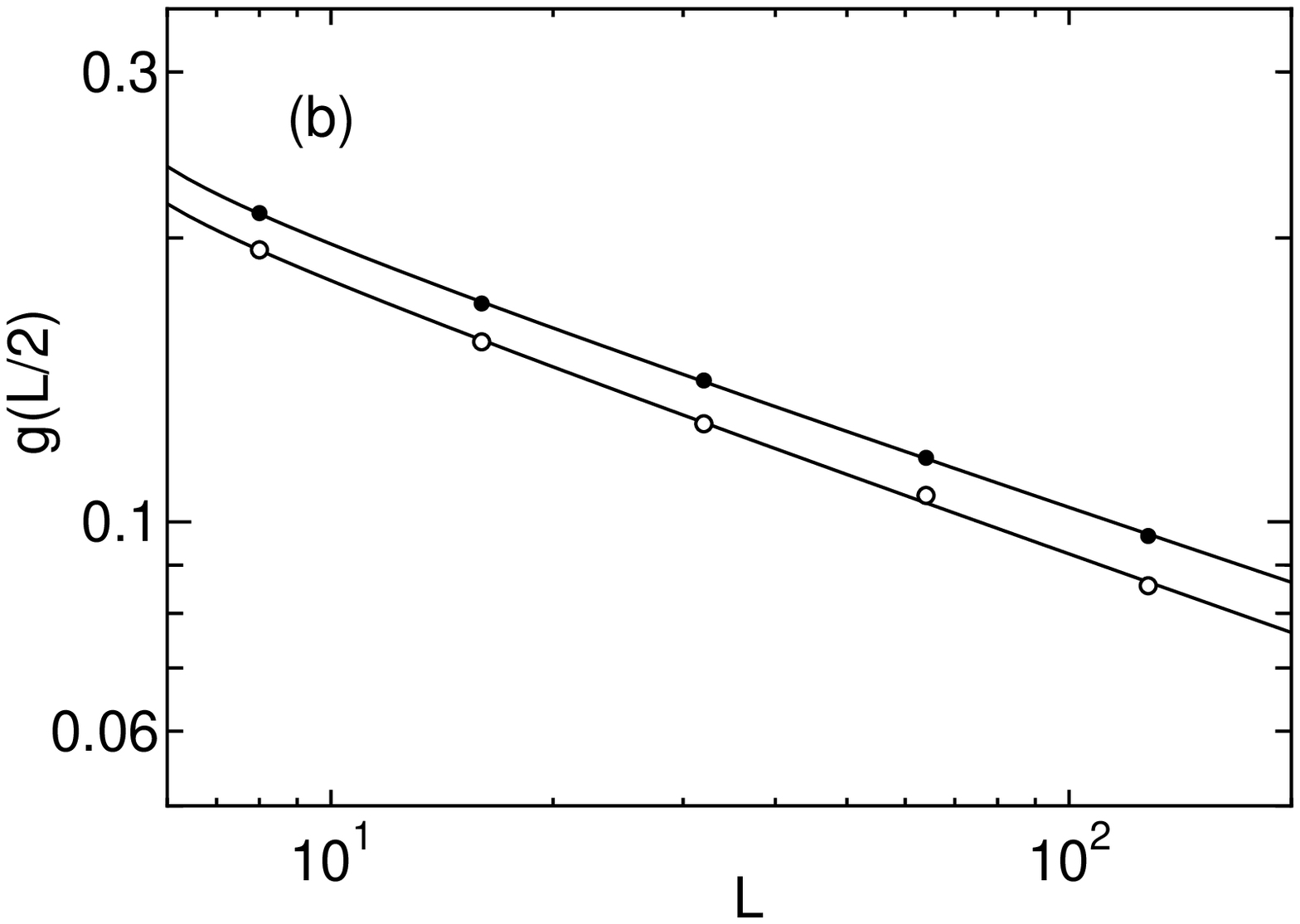}
\end{center}
\caption[]{(a)
Plot of $T_{2}(L)$ and (b) 
the logarithmic plots of $\left< g(L/2) \right>$ 
at $T_2(L)$ 
of the 2D 6-state clock model for $L$ = 
8, 16, 32, 64, and 128, where $l = \ln bL$. 
The closed circles are data for $R_c=0.87$, and 
open circles are those for $R_c=0.85$.
} 
\label{fig_2}
\end{figure}

We use the FSS analysis based on the KT form of 
the correlation length, Eq.~(\ref{corr_length}). 
For the $L$ dependence of $T_{\rm KT}(L)$, 
we have the relation 
\begin{equation}
 T_{\rm KT}(L) = T_{\rm KT} + \frac{c^2 T_{\rm KT}}{(\ln bL)^2}. 
\label{T_KT}
\end{equation}
We plot $T_2(L)$ as a function of $l^{-2}$ 
with $l=\ln bL$ for the best-fitted parameters in Fig.~\ref{fig_2}(a). 
Here, we concentrate on the high-temperature transition $T_2(L)$. 
The value of $R_c$ has been set to be 0.87 and 0.85. 
The error bars are smaller than the size of marks. 
We should mention that the data with different $R_c$ 
are collapsed on a single curve in this plot, which means that 
$b$ depends on $R_c$ in Eq.~(\ref{T_KT}) and the difference of 
$R_c$ can be absorbed in the $R_c$ dependence of $b$. 
The present estimate of $T_2$ is 0.899(3). 
This value is consistent with 
the estimate of the PCC algorithm using the percolating 
properties, 0.9008(6) \cite{XY}.

We also estimate the critical exponent $\eta$ from the size 
dependence of the correlation at $T_2(L)$. 
In Fig.~\ref{fig_2}(b), we plot $\l g(L/2) \r$ at $T_2(L)$ as a function 
of $L$ in a logarithmic scale. 
For the estimate of $\eta$, we use the FSS form 
including small multiplicative logarithmic corrections 
\begin{equation}
 \l g(L/2) \r = AL^{-\eta}(\ln b^{\prime}L)^{-2r}.
\label{eta}
\end{equation}
Our estimate is $\eta$ = 0.250(3) from the data 
for $R_c=0.87$, and $\eta$ = 0.265(2) from the data 
for $R_c=0.85$, which are compatible with the theoretical 
prediction, 1/4 (=0.25).  

The detailed description of the generalized scheme 
of the PCC algorithm based on the FSS of the correlation 
ratio together with its application to the 2D quantum  
XY model of $S=1/2$ will be reported elsewhere \cite{to02}.

\section{Cluster-flip flat histogram method}

In this section, we consider the combination of the cluster algorithm 
and the extended ensemble method.  
One calculates the energy DOS, $g(E)$, 
in the multicanonical method \cite{berg91,Lee93} and 
the Wang-Landau method \cite{wl01}; 
the energy histogram $H(E)$ is checked 
during the Monte Carlo process. 
In contrast, the DOS for bond number $n_b$, 
$\Omega (n_b )$, is calculated in the multibondic ensemble 
method \cite{Janke} or its improvement 
by Yamaguchi and Kawashima \cite{Yama02}; the histogram 
for bond number, $H(n_b)$, is checked in the Monte Carlo 
process.  

In proposing the broad histogram method, 
Oliveira {\it et al.}~\cite{oliveira96} paid attention to 
the number of potential moves, or the number of 
the possible energy change, $N(S, E \to E')$, 
for a given state $S$.  The total number of moves is
\begin{equation}
 \sum_{\Delta E} N(S, E \to E + \Delta E) = N
\end{equation}
for a single-spin flip process, where $N$ is the number of spins. 
The energy DOS is related to the number of potential moves as
\begin{equation}
 g(E) \, \left< N (S, E \to E') \right>_E  
   = g(E') \, \left< N (S', E' \to E) \right>_{E'}, 
\label{BHR}
\end{equation}
where $\left< \cdots \right>_E$ denotes the microcanonical 
average with fixed $E$.  This relation is shown to be valid 
on general grounds \cite{Oliv98,Berg98}, and hereafter we call 
Eq.~(\ref{BHR}) as the broad histogram relation (BHR) 
for the energy. 
One may use the number of potential moves $N(S, E \to E')$ 
for the probability of updating states.  
Alternatively, one may employ other dynamics which has no relation 
to $N(S, E \to E')$, but Eq.~(\ref{BHR}) 
is used when calculating the energy DOS. 

It is interesting to ask whether there is a relation 
similar to the BHR, Eq.~(\ref{BHR}), for the bond number.  
Here, using the cluster (graph) representation, 
we derive the BHR for the bond number.  
We consider the $q$-state Potts model as an example. 
With the framework of the dual algorithm \cite{KD,KG}, 
the partition function is expressed in the double summation 
over state $S$ and graph $G$ as 
\begin{equation}
Z (T) = \sum_{S, G} V_0 (G) \, \Delta (S, G),
\end{equation}
where $\Delta (S, G)$ is a function that takes the value one 
when $S$ is compatible to $G$ and takes the value zero otherwise. 
A graph consists of a set of bonds.  
The weight for graph $G$, $V_0(G)$, is defined as
\begin{equation}
  V_0(G) = V_0(n_b(G),T) = (e^{J/k_BT} - 1)^{n_b(G)} 
\end{equation}
for the $q$-state Potts model, where $n_b(G)$ is 
the number of ``active'' bonds in $G$.
We say a pair $(i, j)$ is satisfied if $\sigma_i = \sigma_j$, 
and unsatisfied otherwise.
Satisfied pairs become active with a probability $p = 1-e^{-J/k_BT}$ 
for given $T$.

By taking the summation over $S$ and $G$ with fixing the number 
of bonds $n_b$, 
the expression for the partition function becomes 
\begin{equation}
Z (T) = \sum_{n_b = 0}^{N_B} \Omega (n_b) \, V_0 (n_b, T), 
\label{eq:d_func_nb}
\end{equation}
where $N_B$ is the total number of nearest-neighbor pairs 
in the whole system.
Here, $\Omega (n_b )$ is the DOS for the bond number defined as
the number of consistent combinations of
graphs and states such that the graph consists of $n_b$ bonds. 
Then, the canonical average of a quantity $A$ is calculated by
\begin{equation}
\left< A \right>_T = 
\frac{\sum_{n_b} \left< A \right>_{n_b} \Omega (n_b ) 
 V_0 (n_b, T)}{Z (T)}, 
\label{eq:canonical}
\end{equation}
where $\left< A \right>_{n_b}$ is 
the microcanonical average with the fixed bond number $n_b$
for the quantity $A$. 
Thus, if we obtain $\Omega(n_b)$ and $\left< \cdots \right>_{n_b}$ 
during the simulation process, we can calculate the canonical 
average of any quantity. 

In deriving the BHR for the bond number,  
we follow a method similar to that used 
to derive the BHR for the bond number \cite{Oliv98,Berg98}. 
Instead of using the relation between states,
we consider the relation between graphs.  
The number of potential moves from the graph 
with the bond number $n_b$ to the graph with $n_b+1$, 
$N(S, G, n_b \to n_b+1)$, for fixed $S$ is equal to 
that of the number of potential moves 
from the graph with $n_b+1$ to that with $n_b$, 
$N(S, G', n_b + 1 \to n_b)$.  
Taking a summation over states $S$ and
using the definition of the microcanonical average 
with the fixed bond number $n_b$, 
we have
\begin{equation}
\Omega (n_b) \left< N(G, n_b \to n_b + 1) \right>_{n_b}
= \Omega (n_b + 1) \, \left< N(G', n_b + 1 \to n_b) \right>_{n_b + 1}.
\label{eq:broad}
\end{equation}
This is the BHR for the bond number.
It should be noted that $N(G, n_b \to n_b + 1)$ is a possible 
number of bonds to add, and related to the number of satisfied pairs 
for the given state $S$, $n_p(S)$, by 
$
 N(G, n_b \to n_b+1) = n_p(S) - n_b. 
$
With use of the microcanonical average with fixed bond number 
for $n_p$, we have the relation 
\begin{equation}
 \left< N(G, n_b \to n_b + 1) \right>_{n_b}
 = \left< n_p \right>_{n_b} - n_b.
\label{eq:transition1}
\end{equation}
On the other hand, the possible number of bonds to delete, 
$N(G', n_b + 1 \to n_b)$, is simply given by $n_b+1$, that is,
\begin{equation}
 \left< N(G', n_b + 1 \to n_b) \right>_{n_b + 1} = n_b + 1.
\label{eq:transition2}
\end{equation}
From the BHR for the bond number, Eq.~(\ref{eq:broad}), we have
\begin{equation}
 \frac{\Omega(n_b)}{\Omega(0)} = 
  \prod_{l=0}^{n_b-1}
  \frac{\left< N(G, l \to l+1) \right>_{n_b=l}}
       {\left< N(G, l+1 \to l) \right>_{n_b=l+1}}
\label{eq:broad2}
\end{equation}
Then, substituting Eqs.~(\ref{eq:transition1}) 
and (\ref{eq:transition2}) into Eq.~(\ref{eq:broad2}), we obtain 
the bond-number DOS, $\Omega (n_b)$, as 
\begin{equation}
\ln \Omega(n_b) = \ln \Omega(0) + \sum_{l=0}^{n_b - 1} \ln
 \biggl(
 \frac{\left< n_p \right>_{n_b=l} - l}{l + 1}.
 \biggr).
\label{eq:Ocal}
\end{equation}
When calculating the bond-number DOS from the BHR for 
the bond number, we only need the information on 
$\left< n_p \right>_{n_b}$, the microcanonical average 
with fixed $n_b$ of the number of satisfied pairs $n_p$. 
It is much simpler than the case of the BHR formulation 
for the energy DOS.  

Moreover, in the computation of $n_p$, we can use 
an improved estimator. If a pair of sites $(i,j)$ belong to 
the different cluster, this pair is satisfied with a probability 
of $1/Q$.  If a pair of sites belong to the same cluster, 
this pair is always satisfied.
Then, we can employ an improved estimator $\tilde{n}_p$ as 
\begin{equation}
\tilde{n}_p(G)  = \biggr( 1 - \frac{1}{Q} \biggl)
 \sum_{\left< i,j \right>} \delta_{c_i(G), c_j(G)} + \frac{N_B}{Q},
\label{eq:np2}
\end{equation}
where $c_i(G)$ represent a cluster that a site $i$ belongs to. 
Only the information on graph is needed.  

Let us consider the update process for the Monte Carlo 
simulation.  
In the multibondic ensemble method, a graph is updated 
by adding or deleting a bond for a satisfied pair 
of sites based on $\Omega(n_b)$ \cite{Janke}. 
We may use the number of potential move for the bond number, 
$\left< N (G, \cdots ) \right>_{n_b}$, for the probability of update. 
Using Eqs.~(\ref{eq:broad}), (\ref{eq:transition1}), and 
(\ref{eq:transition2}), 
we get the probability to delete a bond, 
\begin{equation}
 P(n_b \to n_b - 1) = \frac{\left< n_p \right>_{n_b - 1} + 1 - n_b}
{\left< n_p \right>_{n_b - 1} + 1}, 
\label{eq:prob3}
\end{equation}
and the probability to add a bond, 
\begin{equation}
P(n_b \to n_b + 1) = \frac{n_b + 1}{\left< n_p \right>_{n_b} + 1}, 
\label{eq:prob4}
\end{equation}
respectively.

The actual Monte Carlo procedure is as follows.  
We start from some state (spin configuration) $S$, and 
an arbitrary graph $G$ consistent with it.  We add or delete 
a bond of satisfied pairs with the probability 
(\ref{eq:prob3}) or (\ref{eq:prob4}). 
After making such a process as many as the number of total pairs, $N_B$, 
we flip every cluster 
with the probability 1/2.  And we repeat the process. 
Since we do not know the exact form of $\left< n_p \right>_{n_b}$, 
we use the accumulated average for $\left< n_p \right>_{n_b}$. 
The dynamics proposed here can be regarded as 
the flat histogram method for the bond number, which we call 
the cluster-flip flat histogram method.  
As $\left< n_p \right>_{n_b}$ converges to the exact value, 
the histogram $H(n_b)$ becomes flat.  We calculate 
the bond-number DOS, 
and then calculate various quantities by Eq.~(\ref{eq:canonical}).

\begin{figure}[th]
\begin{center}
\includegraphics[width=.48\textwidth]{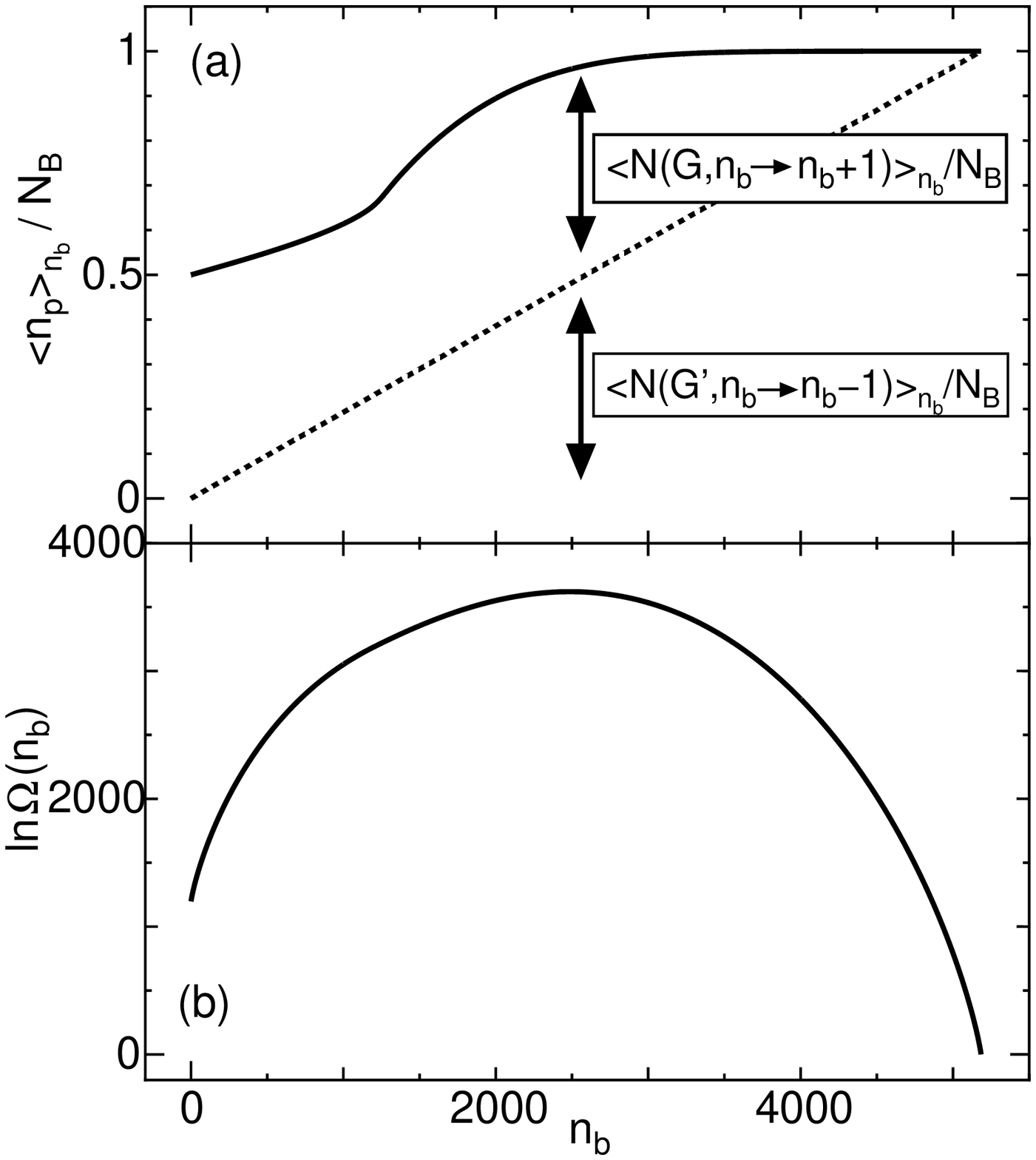}
\hspace{0.02\textwidth}
\includegraphics[width=.48\textwidth]{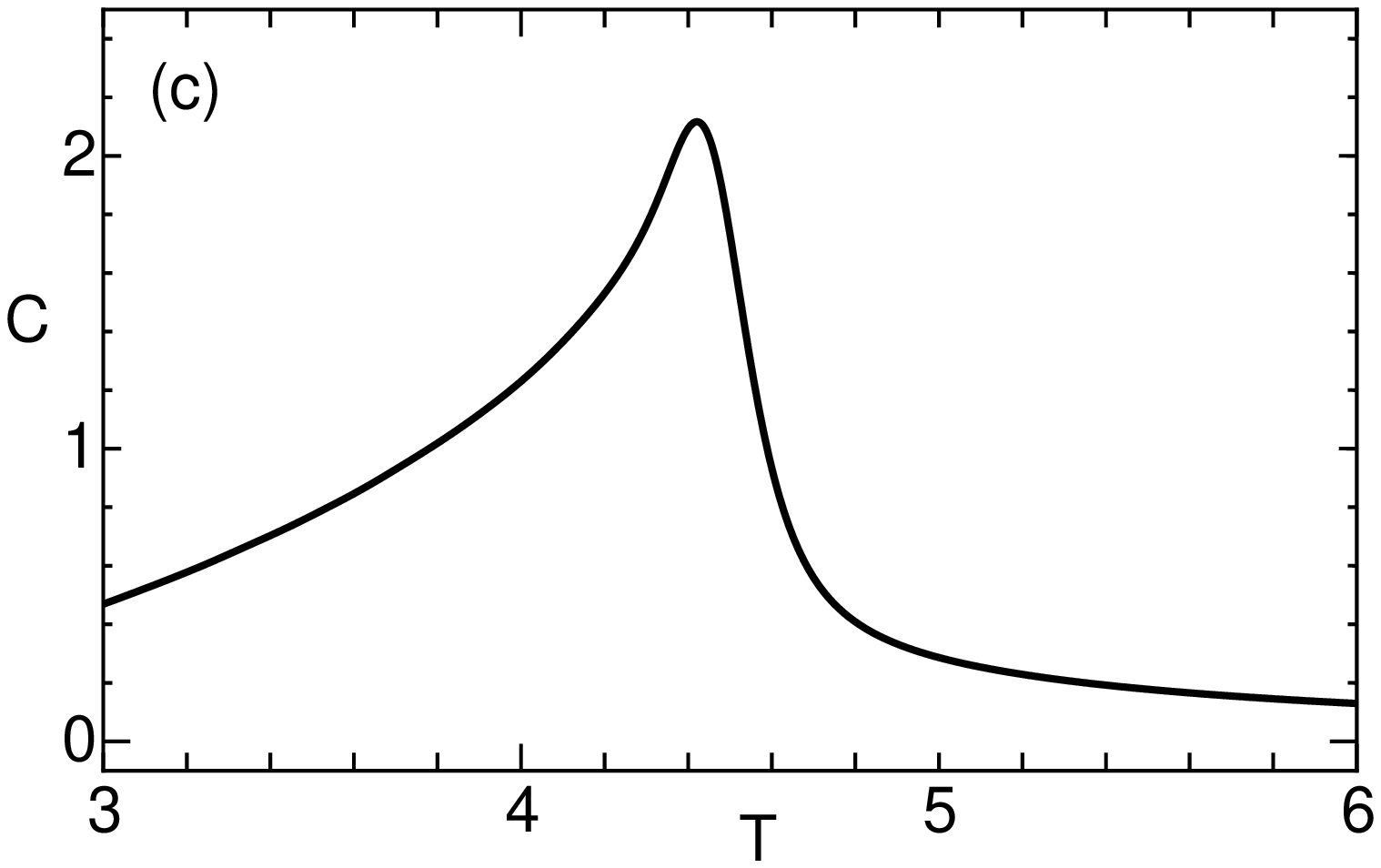}
\end{center}
\caption[]{
(a) $\left< n_p \right>_{n_b}/N_B$ and (b) $\ln \Omega(n_b)$ 
of the $12 \times 12 \times 12$ Ising model 
obtained by the cluster-flip flat histogram method.  
The dotted line in (a) denotes $n_b/N_B$.
(c) The temperature dependence of the specific heat per 
spin, where we have used the units of $J=k_B=1$. }
\label{fig_3}
\end{figure}

As a test, we calculate the $L \times L \times L$ Ising model 
on the simple cubic lattice 
with the periodic boundary conditions by using 
the cluster-flip flat histogram method. 
We show $\left< n_p \right>_{n_b}/N_B$ as a function of $n_b$ for $L=12$ 
by the solid line in Fig.~\ref{fig_3}(a); we give $n_b/N_B$ by the dotted line. 
The number of Monte Carlo sweeps (MCS) is $5 \times 10^7$. 
The difference between the solid and dotted lines represents the number of 
potential moves $\left< N(n_b \to n_b+1) \right>/N_B$, whereas the difference 
between the dotted line and the horizontal axis represents 
$\left< N(n_b \to n_b-1) \right>/N_B$.  We note that 
$\left< n_p \right>_{n_b=0}/N_B = 1/2$, which is expected 
from Eq.~(\ref{eq:np2}). 
The logarithm of the bond-number DOS, $\ln \Omega(n_b)$, obtained 
by $\left< n_p \right>_{n_b}$ is shown in Fig.~\ref{fig_3}(b) as a function of $n_b$.  
The temperature dependence of the specific heat 
is shown in Fig.~\ref{fig_3}(c), which reproduces 
the result obtained by the conventional method. 

\begin{figure}[th]
\begin{center}
\includegraphics[width=.48\textwidth]{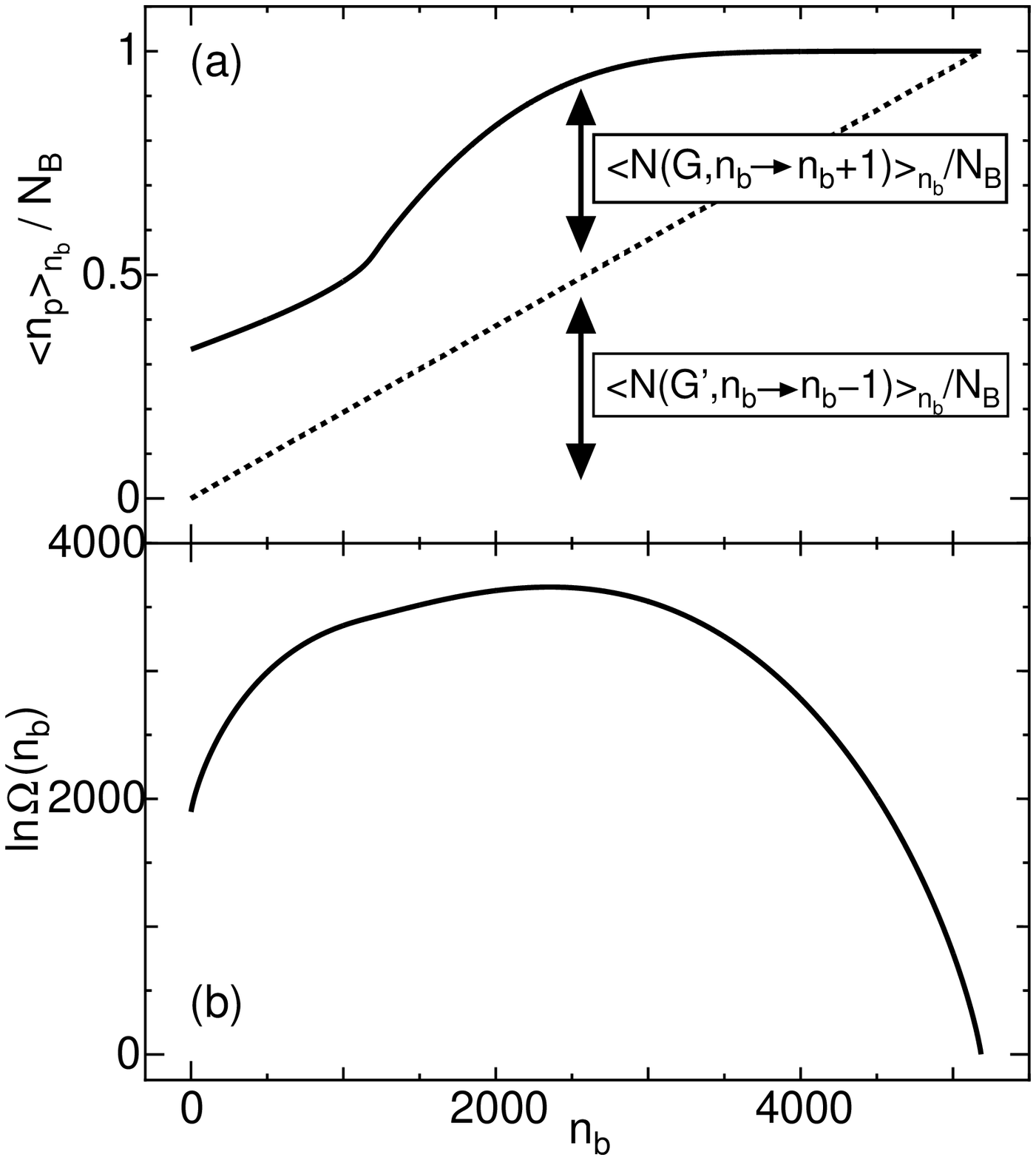}
\hspace{0.02\textwidth}
\includegraphics[width=.48\textwidth]{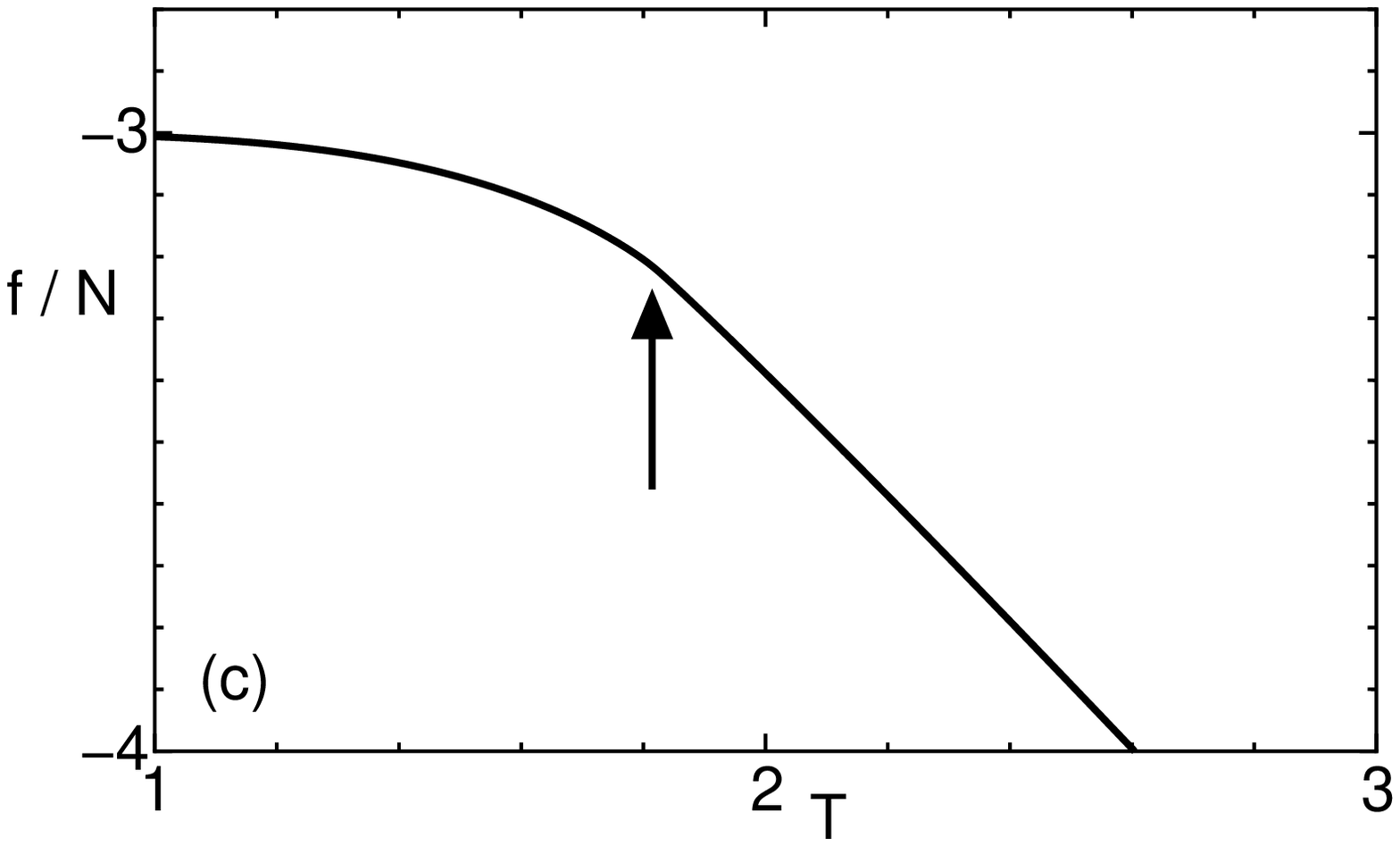}
\end{center}
\caption[]{
(a) $\left< n_p \right>_{n_b}/N_B$ and (b) $\ln \Omega(n_b)$ 
of the 3D 3-state Potts model of $L=12$ 
obtained by the cluster-flip flat histogram method. 
The dotted line in (a) denotes $n_b/N_B$. 
(c) The temperature dependence of the free energy per 
spin, where we have used the units of $J=k_B=1$. 
}
\label{fig_4}
\end{figure}

As another example, we simulate the 3D 3-state Potts model 
on the simple cubic lattice.  A first-order phase transition occurs 
in this model.  
We show $\left< n_p \right>_{n_b}/N_B$ for $L=12$ 
by the solid line in Fig.~\ref{fig_4}(a); we give $n_b$ by the dotted line. 
The number of MCS is $5 \times 10^7$. 
The number of potential moves $\left< N(n_b \to n_b+1) \right>/N_B$ and 
$\left< N(n_b \to n_b-1) \right>/N_B$ are given in the same manner as 
the case of the Ising model.  It is to be noted that 
$\left< n_p \right>_{n_b=0}/N_B = 1/3$ for the 3-state Potts model. 
The logarithm of the bond-number DOS, $\ln \Omega(n_b)$, obtained by 
$\left< n_p \right>_{n_b}$ is shown in Fig.~\ref{fig_4}(b). 
The temperature dependence of the free energy 
is given in Fig.~\ref{fig_4}(c). 
The first-order point where the derivative of the free energy 
has a jump is indicated by the arrow. 

The detailed report on the subject of this section 
will be published in a separate paper \cite{yko02}. 
There, the results for the application to the 2D Ising and 
10-state Potts models will be given, 
and the efficiency of the method will be discussed.

\section{Summary and discussions}
We have discussed the recent progress in Monte Carlo 
algorithms.  First, we have argued the generalization 
of the PCC algorithm based on the study of the FSS 
property of the correlation ratio, 
the ratio of the correlation functions with different distances. 
We apply this generalized scheme of the PCC algorithm to 
the 2D 6-state clock model.  
Since we do {\it not} use the percolating 
property of the system, 
we can apply the PCC algorithm where the mapping 
to the cluster formalism does {\it not} exist. 
It can be applied to many problems.  For example, 
the cluster formalism does not work well for frustrated 
systems, but we can use the generalized PCC algorithm.  
We can also apply the generalized scheme 
to a wide variety of quantum systems. 

Second, we have discussed the combination of the cluster algorithm and 
the extended ensemble method.  
We have derived the rigorous BHR for the bond number, 
investigating the cluster (graph) representation of the spin models.  
We have shown that the bond-number DOS $\Omega(n_b)$ can be calculated 
in terms of $\left< n_p \right>_{n_b}$.   We have proposed 
a Monte Carlo dynamics based on the number of potential moves 
for the bond number, which is regarded 
as the flat histogram method for the bond number.  
We have applied the cluster-flip flat histogram method
to the 3D Ising and 3-state Potts models.  

\section*{Acknowledgments}

We thank N. Kawashima for fruitful discussions and 
the collaboration of a part of the present work.  
We also thank H. Otsuka, J.-S. Wang and C.-K. Hu 
for valuable discussions.
The computation in this work has been done using the facilities of
the Supercomputer Center, Institute for Solid State Physics,
University of Tokyo.
This work was supported by a Grant-in-Aid for Scientific Research 
from the Ministry of Education, Science, Sports and Culture, Japan.


\begin{thebibliography}{8.}
\addcontentsline{toc}{section}{References}

\bibitem{sw87}  R. H. Swendsen and J. S. Wang,
 Phys. Rev. Lett. {\bf 58}, 86 (1987).
\bibitem{wolff89}  U. Wolff,
 Phys. Rev. Lett. {\bf 62}, 361 (1989).
\bibitem{berg91}  B. A. Berg and T. Neuhaus, 
 Phys. Lett. B {\bf 267}, 249 (1991);
 Phys. Rev. Lett. {\bf 68}, 9 (1992).
\bibitem{Lee93}
 J. Lee, Phys. Rev. Lett. {\bf 71}, 211 (1993).

\bibitem{oliveira96} 
 P. M. C. de Oliveira, T. J. P. Penna, and H. J. Herrmann,
 Braz. J. Phys. {\bf 26}, 677 (1996); 
 Eur. Phys. J. B {\bf 1}, 205 (1998).

\bibitem{wang98} J.-S. Wang, 
 Eur. Phys. J. B {\bf 8}, 287 (1998).
\bibitem{Wang00}
 J. S. Wang and L. W. Lee, Comp. Phys. Commun. {\bf 127}, 131 (2000); 
 J. S. Wang, Physica A {\bf 281}, 147 (2000).
\bibitem{wl01} F. Wang and D. P. Landau, 
 Phys. Rev. Lett. {\bf 86}, 2050 (2001);
 Phys. Rev. E {\bf 64}, 056101 (2001).
\bibitem{yo01} 
 C. Yamaguchi and Y. Okabe, 
 J. Phys. A {\bf 34}, 8781 (2001). 

\bibitem{PCC} 
 Y. Tomita and Y. Okabe, 
 Phys. Rev. Lett. {\bf 86}, 572 (2001);  
 J. Phys. Soc. Jpn. {\bf 71}, 1570 (2002).

\bibitem{to01b} 
 Y. Tomita and Y. Okabe, 
 Phys. Rev. E {\bf 64}, 036114 (2001).
\bibitem{XY}
 Y. Tomita and Y. Okabe, Phys. Rev. B {\bf 65}, 184405 (2002).

\bibitem{KT} 
 J. Kosterlitz and D. Thouless, J. Phys. C {\bf 6}, 1181 (1973). 

\bibitem{Janke}
 W. Janke and S. Kappler, Phys. Rev. Lett. {\bf 74}, 212 (1995).
\bibitem{Yama02}
 C. Yamaguchi and N. Kawashima, Phys. Rev. E {\bf 65}, 056710 (2002).

\bibitem{KF}  
 P. W. Kasteleyn and C. M. Fortuin,
 J. Phys. Soc. Jpn. Suppl. {\bf 26}, 11 (1969);
 C. M. Fortuin and P. W. Kasteleyn,
 Physica {\bf 57}, 536 (1972).

\bibitem{Hu84}
 C.-K. Hu, Phys. Rev. B {\bf 29}, 5103 and 5109 (1984). 

\bibitem{Hu92}
 C.-K. Hu, Phys. Rev. B {\bf 46}, 6592 (1992); 
 Phys. Rev. Lett. {\bf 69}, 2739.

\bibitem{Binder} 
  K. Binder, Z. Phys. B {\bf 43}, 119 (1981).

\bibitem{Jose}
 J. V. Jos\'e, L. P. Kadanoff, S. Kirkpatrick, and D. R. Nelson,
 Phys. Rev. B {\bf 16}, 1217 (1977).

\bibitem{to02}
 Y. Tomita and Y. Okabe, to appear in Phys. Rev. B.

\bibitem{Oliv98}
 P. M. C. de Oliveira, Eur. Phys. J. B {\bf 6}, 111 (1998).
\bibitem{Berg98}
 B. A. Berg and U. H. E. Hansmann, Eur. Phys. J. B {\bf 6}, 395 (1998).

\bibitem{KD}
 D. Kandel and E. Domany, Phys. Rev. B {\bf 43}, 8539 (1991).
\bibitem{KG}
 N. Kawashima and J. E. Gubernatis, Phys. Rev. E {\bf 51}, 1547 (1995).  

\bibitem{yko02}
 C. Yamaguchi, N. Kawashima, and Y. Okabe, Phys. Rev. E. {\bf 66}, 
 036704 (2002).

\end{thebibliography}
\end{document}